# Bumblebee field in a Topological Framework

L.P. Colatto*

*CEFET-RJ Campus Petrpolis, CEP24920-003 Rio de Janeiro, RJ, Brasil*

A.L.A. Penna†

*International Center for Condensed Matter Physics, CP 04513,
CEP 70919-970, Universidade de Brasíli, DF, Brasil*

W.C. Santos‡

*Universidade de Brasla, Instituto de Fsica , 70910-900, Braslia, DF, Brasil*
(ΩDated: July 24, 2020)

A vector field coming from spontaneous Lorentz violation mechanism, namely Bumblebee model is analysed in a topological framework in a $(1+2)D$ Minkowski space-time. Taking a $(1+2)D$ nonlinear Bumblebee vector matter field dynamics where we include topological like Chern-Simons type terms, a vector version of a soliton state, or vortex was found. The Nielsen-Olesen procedure was used in order to derive a Lorentz-violation vector parameter which characterizes, via Spontaneous Symmetry Breaking mechanism, the non-trivial vacuum. We verify the stability of the model as much as the magnetic vortex, and noticed that the soliton modes with polarized direction generated can be associated with local anisotropy of vacuum energy. The vortex equations of motion and the asymptotic behaviour is presented. We have obtained that the effect of the Lorentz symmetry violation expressed by the a time-like Bumblebee vector field vacuum could be shown as kind of pulse at a fixed point $r_0$ in a limitless universe, or as a barrier at $r_0$ which can represent a boundary in the universe, if the Bumblebee vector field vacuum has space-like characteristic. We also analyse the spectrum via propagators where we note that the topological mass contributes as well to the dynamical mass poles. We obtain that the Chern-Simons type terms, in fact, indicates the "speed" of the field to saturate the asymptotic limit and that the vortex core can not be dimension zero.

## I. INTRODUCTION

According to the Standard Model, the spontaneous violation of symmetry and its consequences are well established. The existence of nontrivial vacuum expectation value directly modifies the properties of fields that couple to it and can indirectly modify them through interactions with other affected fields. The vacuum value $\langle\phi\rangle$ of the Higgs field spontaneously breaks the $SU(2) \times U(1)$ gauge symmetry associated with the electro-weak force and generates masses for several particles, and separates the electromagnetic and weak forces. The W and Z bosons are the elementary particles that mediate the weak interaction, while the photon mediates the electromagnetic interaction. At energies much greater than 100 GeV all these particles behave in a similar manner. The Weinberg-Salam theory predicts that, at lower energies, this symmetry is broken so that the photon and the massive W and Z bosons emerge . As we are dealing with non-linear models, it is necessary to include soliton-type configuration in order to achieve a close to actual physical states, which stability is highly dependent on the topological characteristic of the boundary conditions [1, 2].

On the other hand, some time ago, from the point of view of the vacuum at the Planck scale of the string theory, it was observed a possibility of a violation (or breaking) of Lorentz symmetry, similar to the idea of a spontaneous symmetry breaking (SSB) via mechanism of Higgs field. The spontaneous Lorentz Symmetry violation with aid of a vector field, called Bumblebee vector field, has been extensively studied from pioneer works of V. A. Kosteleck and S. Samuel [3] and consequently there was an increasing number of works on the theme of Lorentz symmetry violation via spontaneous symmetry breaking of a vector field [4–8]. Remarkably the Bumblebee field does not obey $U(1)$ gauge symmetry, as a matter field, and acquires a non-zero vector vacuum expectation value inducing a spontaneous Lorentz symmetry violation.

In fact studies were done on quantum gravity theories predicting the possibility of Lorentz symmetry violation via the Bumblebee field [3]. For instance, a Bumblebee field $B_\mu$ coupled to gravity, with generalized quadratic kinetic terms involving up to second-order derivatives in $B_\mu$, and with an Einstein-Hilbert term in Riemann space-time, is given by the Lagrangian density [6]:

$$\mathcal{L}_B = \sqrt{-g}\left[\frac{1}{2\kappa}(R-2\Lambda) + \frac{1}{2\kappa}\xi B^\mu B^\nu R_{\mu\nu} + \rho B^\mu B_\mu R + \frac{\sigma}{2}(\nabla_\nu B_\nu)(\nabla^\mu B^\nu) + \frac{\tau}{2}(\nabla_\mu B^\mu)(\nabla_\nu B^\nu) + \right.$$
$$\left. -\frac{1}{4}B_{\mu\nu}B^{\mu\nu} - V(B_\mu B^\mu)\right], \tag{1}$$

where $R$ is the curvature scalar, $R_{\mu\nu}$ is the Ricci tensor, $\kappa = 16\pi G$ and $\xi$, $\rho$, $\sigma$ and $\tau$ are parameters. The Bumblebee field strength is defined as

$$B_{\mu\nu} = \nabla_\mu B_\nu - \nabla_\nu B_\mu = \partial_\mu B_\nu - \partial_\nu B_\mu. \tag{2}$$

---

*Electronic address: lcolatto@gmail.com
†Electronic address: penna.andre@gmail.com
‡Electronic address: wytler@fis.unb.br

The potential $V$ providing a non-zero vacuum expectation value for vector field $B_\mu$ has the following function form

$$V = -\alpha^2 B_\mu B^\mu - \lambda(B_\mu B^\mu)^2. \tag{3}$$

Then, the non-zero vacuum expectation value of the Bumblebee field is obtained by

$$\frac{\partial V}{\partial B_\mu} = -\alpha^2 B^\mu - 2\lambda(B_\nu B^\nu)B^\mu = 0 \tag{4}$$

Therefore it is solved when the field $B^\mu$ acquires a non-zero vacuum expectation value,

$$\langle B^\mu \rangle = b^\mu, \tag{5}$$

where $b^\mu$ has constant magnitude but different orientation at different space-time points, and therefore the global Lorentz and translation symmetries are broken [5]. It is remarkable that the potential has a minimum with respect to its argument or is constrained to zero when

$$B^\mu B_\mu = \mp|b|^2. \tag{6}$$

So, we state that the magnitude of the Bumblebee vector field is a time-like vector when we have $+|b|^2$ and the Bumblebee vector field is a space-like vector when we have $-|b|^2$.

Indeed since the birth of quantum mechanics there has been a growing and continuing interest in the study of possible Lorentz violation effects and its possible origins as the non-commutative phenomena [9–20]. In the context of non-commutative quantum field theory, deformation terms appear due to a non-commutative spcetime algebra applied to quantum field theory what can include terms that violate Lorentz invariance due to the preferred direction of non-commutativity. So, a deformation parameter can be introduced in algebras reaching to quantum groups, which develop an interesting theoretical laboratory to extend the Lorentz symmetry framework [21]. Returning to Lorentz symmetry violation issue, Colladay and Kostelecký have observed the possibility to implement Lorentz violating extensions to the Standard Model that was extensively investigated as a reflection issue of the possible Planck scale original scenario [22, 23]. Remarkably many effects are conjectured to have a Lorentz symmetry violation origin, e.g., some astrophysical phenomena could be fitted by these extensions [24], and theoretical developments have been largely worked on Ref. [25–27].

On the other hand, one important framework to study topological effects are present in $(1+2)D$ models. Where these effects can be theoretically represented through the vacuum of a $(1+2)D$ model with Chern-Simons photon coupling, what is called Carroll-Jackiw field model [28]. That work has presented a preferred space time direction dictated by the introduction of a constant four-vector $b_\mu$. Consequently, the photon circular polarization shows a non usual dispersion relation [29], while the linear polarized photons exhibit optical birefringence. Furthermore the four-vector induces anisotropic optical vacuum which could be detected through the observation of tiny Lorentz symmetry violation effects ($b_\mu b^\mu < 10^{-33}eV \simeq 10^{-28}cm^{-1}$) on space-time [30–33]. In a previous work, we suggest that LSB terms, with consequent anisotropic optical effects, could originate from a spontaneous symmetry breaking (SSB) mechanism on a matter vector field [34]. Taking this into account, the reduced planar surface model to $(1+2)D$ may reveal the contribution of topological objects to such non-trivial optical vacuum effects. Our purpose in this work is to obtain stable $(1+2)D$ vortex lines solutions, starting from interacting of Bumblebee matter vector field with the $U(1)$ gauge vector field. We will discuss the contribution of the Chern-Simons-type terms to asymptotic behaviour of the Bumblebee matter field.

The outline of the work is the following: In Sec. II we consider the Bumblebee matter vector field model and we discuss how it can arise vortices solutions. Sec. III is devoted to analyse the asymptotic physical solutions and the Bumblebee mass contribution to the vortex lines and the stability of the model. In Sec. IV we study of the mass poles through the propagators obtained and discuss the relations to the vacuum and asymptotic limit of the fields. Finally, in the Conclusion we discuss the results and give perspectives of new approaches to the subject

## II. THE BUMBLEBEE FIELD VORTEX

It is worth to remark that Bumblebee models are not gauge invariant [6, 8], for this reason we propose to assume that the Bumblebee field plays the role of a matter field in Minkowski space-time. Furthermore we add the condition that Bumblebee field can be electrically charged, i.e., we assume a charged complex Bumblebee vector field that maintain the global $U(1)$ invariance. For our proposals we are going to take the parameters $\xi$, $\rho$ and $\sigma$ of the equation (1) to be null, as well as $g_{\mu\nu} = \eta_{\mu\nu} = \mathrm{diag}(+ - -)$ with $\sqrt{g} = 1$, so that the Lagrangian (1) is modified as follows,

$$\mathcal{L}_B = -\frac{1}{2} B^*_{\mu\nu} B^{\mu\nu} - (\partial_\mu B^\mu)^* (\partial_\nu B^\nu) + \alpha^2 B^*_\mu B^\mu + \lambda (B^*_\mu B^\mu)^2. \tag{7}$$

Taking a previous work, [34], we use the charged matter vector field model and we assume it is living in a $(1+2)$D world, including global $U(1)$ invariant terms, and topological Chern-Simons-type terms. It can be written as,

$$\mathcal{L} = -\frac{1}{2} B^*_{\mu\nu} B^{\mu\nu} - (\partial_\mu B^\mu)^* (\partial_\nu B^\nu) + m\epsilon^{\mu\nu\kappa} B_\mu \partial_\nu B^*_\kappa + m\epsilon^{\mu\nu\kappa} B^*_\mu \partial_\nu B_\kappa + \alpha^2 B^*_\mu B^\mu + \lambda (B^*_\mu B^\mu)^2, \tag{8}$$

where $B_\mu$ is a Bumblebee charged vector field and $B_{\mu\nu} = \partial_\mu B_\nu - \partial_\nu B_\mu$ is the Bumblebee field strength. The symmetry $U(1)$ allows us to insert a topological massive Chern-Simons-type term that does not play a role in the vacuum achievement but will be important to the mass term definition in the asymptotic limit behaviour that will be made clear further ahead[1]. As we are treating $B_\mu$ as a matter field the global $U(1)$ symmetry induces the presence of a dynamical longitudinal term in the Lagrangian (8). We can verify that this model gives rise to a SSB mechanism whose non-trivial vacuum [25, 26, 34] allows a particular soliton solution. Analogously to the usual scalar case we are going to assume asymptotic values on the boundary, namely $S^1$ topological circle recalling the Nielsen-Olesen-type approach. Consequently, taking $r$ and $\theta$ as the polar coordinates, we construct the following values for $B_\mu$ at $r \to \infty$,

$$B_\mu = b_\mu e^{in\theta} \qquad \text{and} \qquad B^*_\mu B^\mu = b_\mu b^\mu = \pm |b|^2 \tag{9}$$

where $b_\mu$ is a constant vector that is defined by a minimal value of the energy of the Lagrangian (8), as it also has been assumed in [3–8, 34]. The result is that the Lagrangian (8) allows a SSB on Bumblebee matter vector field $B_\mu$ in accordance with the equation (4), where we have,

$$b_\mu b^\mu = \frac{-\alpha^2}{2\lambda}. \tag{10}$$

So, we can specify two directions in space-time for the above vector, $b_\mu$: time-like and space-like. The unitary vector $u_\mu$ in the $(1+2)D$ Minkowski space-time, has the property that $u_\mu u^\mu = \pm 1$. We state that:

- *Time-like* $b_\mu$, where we have $b_\mu b^\mu = +|b|^2 > 0$, and $b_\mu$ has the same orientation of unitary vector $u_\mu$ when $\alpha^2 < 0$ and $u_\mu u^\mu = +1$. Thus, the $b_\mu$ is given by,

$$b_\mu = \sqrt{\frac{-\alpha^2}{2\lambda}} \, u_\mu \,. \tag{11}$$

- *Space-like* $b_\mu$, where we have $b_\mu b^\mu = -|b|^2 < 0$, and $b_\mu$ has the same orientation of unitary vector $u_\mu$ when $\alpha^2 > 0$ and $u_\mu u^\mu = -1$. Thus, the $b_\mu$ must be given by,

$$b_\mu = \sqrt{\frac{+\alpha^2}{2\lambda}} \, u_\mu \,, \tag{12}$$

where it should be noted that the magnitude $b_\mu b^\mu$ in the equation (10) is retrieved in both the cases.

Another possibility to the $b_\mu$ is to be a light-like vector, where $b_\mu b^\mu = 0$. However, due to the equation (10), this possibility cancels the spontaneous symmetry breaking.

Hence our particular non trivial choices for $B_\mu$ at infinity are there that can lead us to fixed and exact values for the degenerated vacua. Emphasizing that the important aspect of the SSB mechanism applied to the Bumblebee matter vector field is that it generates Lorentz-violation parameters spontaneously [34].

---

[1] The two Chern-Simons-type terms are indeed symmetric, we have written them to observe the global $U(1)$-symmetry explicitly.

The vortex lines structures can be derived from the non-trivial topology of the vacuum of the Bumblebee matter vector field. From the anisotropic vacuum in $(1+2)D$, the time-component $(\mu = 0)$ and the space-component $(\mu = i)$ are written as,

$$B_0 = b_0 e^{in\theta}, \qquad\qquad B_i = b_i e^{in\theta}, \tag{13}$$

where $B_i$ assumes two components $B_r$ and $B_\theta$ in the polar plane as follows,

$$B_r = b_r e^{in\theta}, \qquad\qquad B_\theta = b_\theta e^{in\theta} \tag{14}$$

where $b_0$ is a time component and $b_i$ ($b_r$ and $b_\theta$) are space (radial and angular) vector components of the Lorentz-violating parameter $b_\mu$. Many recent discussions related to the existence of these parameters have been considered in Ref. [25, 26, 32]. In our case, the $B_\mu$ field generates a vortex whose stability is analysed starting from static configuration for the Hamiltonian functional density,

$$\mathcal{H}_{s.c.} = -(\vec{\nabla} B_0)^* \cdot (\vec{\nabla} B_0) + (\vec{\nabla} \times \vec{B})^* \cdot (\vec{\nabla} \times \vec{B}) + (\vec{\nabla} \cdot \vec{B})^* \cdot (\vec{\nabla} \cdot \vec{B}), \tag{15}$$

where we have also admitted a static configuration for the SSB potential,

$$V(B_\mu, B_\mu^*) = \lambda[b^2 - B_\mu^* B^\mu]^2 = 0, \tag{16}$$

which results precisely in the expression (9) on the boundary. Let us now treat the system (15) in three dimensions, admitting a cylindrical or axial symmetry around the $z$-axis. Thus, the $z$-component of $B_\mu$ is assumed to be constant, and from (9) we have found that,

$$\vec{\nabla} B_0 = \frac{inb_0}{r} e^{in\theta} \hat{\theta}, \qquad \vec{\nabla} \cdot \vec{B} = \frac{1}{r}(b_r + inb_\theta) e^{in\theta}, \qquad \vec{\nabla} \times \vec{B} = \frac{1}{r}(b_\theta - inb_r) e^{in\theta} \hat{z}. \tag{17}$$

Hence substituting (17) with (15) we find that at $r \to \infty$ the Hamiltonian functional density is given by,

$$\mathcal{H}_{s.c.} = \frac{n^2 b_0^2 + (n^2+1)(b_r^2 + b_\theta^2)}{r^2} \tag{18}$$

and the energy of the vortex relative to the anisotropic vacuum parameters is,

$$E = \int^\infty \mathcal{H}_{s.c.} \, r dr d\theta = \left[ n^2 b_0^2 + (n^2+1)\left(b_r^2 + b_\theta^2\right) \right] \ln |r| \Big|^\infty \tag{19}$$

As the scalar case, we treat this logarithmic divergence by adding a gauge field $A_\mu$ in the model, so,

$$D_\mu B_\nu = \partial_\mu B_\nu + ieA_\mu B_\nu. \tag{20}$$

We shall then assume the gauge choice $A_0 = 0$, where $A^\mu = \vec{A} = A^r \hat{r} + A^\theta \hat{\theta} = \frac{1}{e} \vec{\nabla}(n\theta)$, for very large $r$ value. As is well known, we write it analytically in cylindrical symmetry, or

$$A_r \to 0 \qquad\qquad \text{and} \qquad\qquad -A^\theta = A_\theta, \to -\frac{n}{er} \tag{21}$$

since we treat the covariant derivative (20) at infinity, the results is

$$D_\mu B_\nu = 0. \tag{22}$$

Thus, the stability of this model requires a "vector electrodynamics" whose Lagrangian is defined by,

$$\mathcal{L} = -\frac{1}{4} f_{\mu\nu} f^{\mu\nu} - \frac{1}{2} \mathcal{B}_{\mu\nu}^* \mathcal{B}^{\mu\nu} - (D_\mu B^\mu)^*(D_\nu B^\nu) + m\epsilon^{\mu\nu\kappa} B_\mu^*(D_\nu B_\kappa) + m\epsilon^{\mu\nu\kappa} B_\mu(D_\nu B_\kappa)^* + \alpha^2 B_\mu^* B^\mu + \lambda(B_\mu^* B^\mu)^2, \tag{23}$$

where $\mathcal{B}_{\mu\nu} = D_\mu B_\nu - D_\nu B_\mu$ and $f_{\mu\nu} = \partial_\mu A_\nu - \partial_\nu A_\mu$. From the relations (21) which define a pure gauge, the Lagrangian model (23) assumes a finite energy configuration for its Hamiltonian functional density at infinity,

$$\mathcal{B}_{\mu\nu} \to 0 \qquad\qquad \text{and} \qquad\qquad \mathcal{H} \to 0 \tag{24}$$

and the gauge field $A_\mu$ gives rise to a soliton magnetic flux or vortex,

$$\Phi = \oint \vec{A} \cdot d\vec{l} = \frac{2\pi n}{e}, \tag{25}$$

which is analogous to the scalar field model and represents a magnetic flux quantization as in the London and Ginzburg-Landau equations to describe the type-II superconductors [1].

## III.  TOPOLOGICAL MASS AND THE BUMBLEBEE FIELD VORTEX

Based on the Lagrangian (23), we now analyse the emergence of soliton kind solutions based on the Nielsen-Olesen work [35], namely vortices, with topological mass contribution of the Bumblebee field. In fact, it yields a vortex solution whose dynamical equation for $A_\mu$ is written as

$$ie(B^*_\nu B^{\mu\nu} - B_\nu B^{\mu\nu *}) + ie(B^{\mu *}\partial_\nu B^\nu - B^\mu \partial_\nu B^{\nu *}) - iem\epsilon^{\mu\nu\kappa}(B^*_\nu B_\kappa - B_\kappa B^*_\nu) - 2e^2 A^\mu (B^*_\nu B^\nu) = \partial_\nu f^{\mu\nu}, \qquad (26)$$

and the dynamical equation for $B_\mu$ is given by

$$D_\mu D^\mu B^\nu + 2m\epsilon^{\nu\kappa\lambda} D_\kappa B_\lambda = -\left[\alpha^2 + 2\lambda(B^*_\mu B^\mu)\right] B^\nu. \qquad (27)$$

We are seeking for vortex solutions in the system described by (26). In order to extract further information about the system we assume cylindrical coordinates in such way that

$$A^\mu = (0, 0, A^\theta) = (0, 0, A(r)) \qquad\text{and}\qquad B_\mu = \beta_\mu(r)e^{in\theta}, \qquad (28)$$

where the asymptotic behaviours of $\beta_\mu(r)$ are given by,

$$\lim_{r\to\infty}\beta_\mu(r) = b_\mu \qquad\qquad \lim_{r\to 0}\beta(r) = 0, \qquad (29)$$

with $\mu = 0,\, r,\, \theta$ which are, respectively, the time, $r$ and $\theta$ cylindrical components of the Bumblebee field. So $B^*_\mu B^\mu = \beta_\mu(r)\beta^\mu(r) = \beta(r)^2$ in such way that at infinity $\beta(r)^2 \longrightarrow (\beta(r\to\infty))^2 = b^2$. In the Minkowski space $(1+2)D$ we can observe that,

$$\beta(r)^2 = \beta_0(r)^2 - \beta_r(r)^2 - \beta_\theta(r)^2 \qquad\text{and so}\qquad \lim_{r\to\infty}\beta(r)^2 = b_0^2 - b_r^2 - b_\theta^2 = b^2. \qquad (30)$$

$|b|$ represents the parameter of Lorentz symmetry violation. Note that the magnitude of vector $\beta^\mu(r)$ can have positive magnitude, time-like vector, or negative magnitude, a space-like vector. As the gauge field $A^\mu$ is a function that exclusively depends on the variable $r$ the equations of motion (26) assumes, in the static configuration, the form:

$$\partial_i f^{\mu i} = ie[B^*_\nu B^{\mu\nu} - B_\nu(B^{\mu\nu})^* + (B^\mu)^*(\partial_\nu B^\nu) - B^\mu(\partial_\nu B^\nu)^*] - 2e^2 A^\mu B^*_\mu B^\nu. \qquad (31)$$

The assumption (28) implies to $\epsilon^{\mu\nu\kappa}(B^*_\nu B_\kappa - B_\kappa B^*_\nu) = 0$, and the equation of motion (27) results in

$$D_i D^i B^\mu + 2m\epsilon^{\mu\kappa\lambda} D_\kappa B_\lambda = -\left[\alpha^2 + 2\lambda(B^*_\nu B^\nu)\right] B^\mu. \qquad (32)$$

Moreover the assumption (28) indicates that the only non null component of $A^\mu$ vector is $A^\theta$, so the term $\partial_i f^{\theta i}$ reduces to $\partial_r f^{\theta r} = \partial_r(\partial^\theta A^r - \partial^r A^\theta) = \partial_r(\nabla\times\vec{A})_k$. The equation of motion (31) for gauge field $A(r)$ can be written down as,

$$r^2\frac{d^2 A(r)}{dr} + r\frac{dA(r)}{dr} + \left[2e^2|\beta(r)|^2 r^2 - 1\right]A(r) = 2en|\beta(r)|^2 r, \qquad (33)$$

and the equation of motion (32) for Bumblebee field yields three equations,

$$\frac{1}{r}\frac{d}{dr}\left[r\frac{d\beta_0(r)}{dr}\right] - \left[\left(\frac{n}{r} - eA(r)\right)^2 + \alpha^2 + 2\lambda|\beta(r)|^2\right]\beta_0(r) = 2m\left[\frac{d\beta_\theta(r)}{dr} - \left(\frac{in}{r} + ieA(r)\right)\beta_r(r)\right], \qquad (34)$$

$$\frac{1}{r}\frac{d}{dr}\left[r\frac{d\beta_r(r)}{dr}\right] - \left[\left(\frac{n}{r} - eA(r)\right)^2 + \alpha^2 + 2\lambda|\beta(r)|^2\right]\beta_r(r) = -2m\left[\frac{in}{r} + ieA(r)\right]\beta_0(r), \qquad (35)$$

$$\frac{1}{r}\frac{d}{dr}\left[r\frac{d\beta_\theta(r)}{dr}\right] - \left[\left(\frac{n}{r} - eA(r)\right)^2 + \alpha^2 + 2\lambda|\beta(r)|^2\right]\beta_\theta(r) = 2m\frac{d\beta_0(r)}{dr}. \qquad (36)$$

Unfortunately, no exact solution can be obtained for these four equations. However, it is possible to establish some important approximate asymptotic solutions for the following cases below.

### A. Vortex vacuum solution

Taking the equation (33), one simple solution can be obtained by using the vacuum limit at large $r$ condition $\beta(r)^2 = b^2$ constant as expressed in equation (30). We must introduce the conditions of time-like Bumblebee vector field magnitude, where $b^2 > 0$ and space-like Bumblebee vector magnitude where $b^2 < 0$. For $b^2 > 0$, where $b^2 = +|b|$ it is straightforward to see that the equation (33) is

$$r^2 \frac{d^2 A(r)}{dr} + r \frac{dA(r)}{dr} + \left[2e^2|b|^2 r^2 - 1\right] A(r) = 2en|b|^2 r \,. \tag{37}$$

The above equation is slightly different from the usual vacuum solution of the Nielsen-Olesen scalar case [1, 35]. One can obtain the potential electromagnetic field $A(r)$ and calculate the $z$-component magnetic field $H(r)$ as,

$$H(r) \equiv \frac{1}{r} \frac{d\left[rA(r)\right]}{dr} \,. \tag{38}$$

We obtain the following solutions to the fields as functions on the Lorentz violation parameter $|b|$,

$$A(r) \;=\; \frac{n}{er} + \frac{C}{e} J_1\!\left(\sqrt{2}|eb|r\right) \quad \overrightarrow{r \to \infty} \quad \frac{n}{er} + \frac{C}{e} \sqrt{\frac{1}{\pi\sqrt{2}|eb|r}} \;\left[\sin(\sqrt{2}|eb|r) - \cos(\sqrt{2}|eb|r)\right] \,, \tag{39}$$

$$H(r) \;=\; C\sqrt{2}\,|b|\, J_0\!\left(\sqrt{2}|eb|r\right) \quad \overrightarrow{r \to \infty} \quad C\sqrt{2}\,|b|\, \sqrt{\frac{1}{\pi\sqrt{2}|eb|r}} \;\left[\sin(\sqrt{2}|eb|r) + \cos(\sqrt{2}|eb|r)\right] \,, \tag{40}$$

where $J_0\!\left(e|b|r\right)$ and $J_1\!\left(e|b|r\right)$ are the Bessel functions and $C$ is a constant. Note that the fields $A(r)$ and $H(r)$ depend on the symmetry violation parameter $|b|$. Moreover, we observe that, in these conditions, the topological mass does not contribute to the electromagnetic potential vector and, consequently, to the magnetic field.

For $b^2 < 0$, where we write $b^2 = -|b|$, the equation (33) becomes,

$$r^2 \frac{d^2 A(r)}{dr} + r \frac{dA(r)}{dr} - \left[2e^2|b|^2 r^2 + 1\right] A(r) = -2en|b|^2 r \,. \tag{41}$$

In this case, where the Bumblebee field is space-like vector, the electromagnetic potential vector and magnetic field are similar to the Nielsen-Olesen scalar case [1, 35], where we have,

$$A(r) \;=\; \frac{n}{er} + \frac{C}{e} K_1\!\left(\sqrt{2}|eb|r\right) \quad \overrightarrow{r \to \infty} \quad \frac{n}{er} + \frac{C}{e} \sqrt{\frac{\pi}{2\sqrt{2}|eb|r}} \; e^{-\sqrt{2}|eb|r} \,, \tag{42}$$

$$H(r) \;=\; C\sqrt{2}\,|b|\, K_0\!\left(\sqrt{2}|eb|r\right) \quad \overrightarrow{r \to \infty} \quad C\sqrt{2}\,|b|\, \sqrt{\frac{\pi}{2\sqrt{2}|eb|r}} \; e^{-\sqrt{2}|eb|r} \,, \tag{43}$$

where $K_0$ and $K_1$ are modified Bessel functions. Another possible solution is taking the magnitude of the Bumblebee vector field be zero or $b^2 = 0$, but this solution is disregarded because $|b| = 0$ imposes the nullity of the spontaneous Lorentz violation.

### B. Vortex core solutions

We are now looking for some approximate solution dealing with topological mass for the Bumblebee fields components $\beta_0(r)$, $\beta_r(r)$, $\beta_\theta(r)$. To this goal it requires to focus on solutions in eqs. (34), (35) and (36) with topological mass at $r < \infty$. It can be done by eliminating the factors $\left[\left(\frac{n}{r} - eA(r)\right)^2 + \alpha^2 + 2\lambda|\beta(r)|^2\right]$ and $\left(\frac{in}{r} + ieA(r)\right)$, so we obtain that

$$\frac{1}{r} \frac{d}{dr} \left[r \frac{d\beta_0(r)}{dr}\right] - \frac{\beta_r}{r\beta_0} \frac{d}{dr} \left[r \frac{d\beta_r(r)}{dr}\right] + \chi \left(\frac{\beta_r^2}{\beta_0} - \beta_0\right) = 2m \frac{d\beta_\theta(r)}{dr} \,, \tag{44}$$

with

$$\chi = \frac{2m}{\beta_\theta} \frac{d\beta_0(r)}{dr} - \frac{1}{r\beta_\theta} \frac{d}{dr} \left[r \frac{d\beta_\theta(r)}{dr}\right] \,. \tag{45}$$

This equation exhibits the following conditions to the topological mass at $r < \infty$ :

1. If simultaneous $\beta_0(r) = constant$ and $\beta_\theta(r) = constant$ it cancels out the topological mass.

2. If $\beta_r(r) = 0$, it does not cancel the topological mass.

3. If $\beta_0(r)$ or $\beta_\theta(r)$ are not simultaneously constant it does not cancel the topological mass.

So, based on the condition 3, we have two possibilities of the simplified equations (44).

The first possibility is obtained if we consider $\beta_r(r) = 0$ and $\beta_0(r) = b_0 = constant$. It results that

$$\frac{d^2\beta_\theta}{dr^2} + \left(\frac{1}{r} - \frac{2m\beta_\theta}{b_0}\right)\frac{d\beta_\theta}{dr} = 0, \tag{46}$$

where the topological mass explicitly depends on the angular component of the Bumblebee field. Note that, despite the very simplified equation obtained, this equation is non linear (46) and has no exact solution. We can go to a step further, considering regions where $\frac{1}{r} \ll \frac{2m\beta_\theta}{b_0}$, such that $1/r$ is negligible, which leads (46) to the equation:

$$\frac{d}{dr}\left(\frac{d\beta_\theta}{dr} - \frac{m}{b_0}\beta_\theta^2\right) = 0\,. \tag{47}$$

The second possibility is obtained if we consider $\beta_r(r) = 0$ and $\beta_\theta(r) = b_\theta = constant$, in this case we get the following equation

$$\frac{d}{dr}\left(\frac{d\beta_0}{dr} - \frac{m}{b_\theta}\beta_0^2\right) = 0\,, \tag{48}$$

where the topological mass explicitly depends on the time component of the Bumblebee field. We observe the similarity of the equations (47) and (48). Using the equation (47) as a guide and taking its analytical solution, we have $\left(\frac{d\beta_\theta}{dr} - \frac{m}{b_0}\beta_\theta^2\right) = -c_1$, with $c_1$ a positive constant [2], we get

$$\beta_\theta = -\sqrt{\frac{b_0\,c_1}{m}}\tanh\left[\sqrt{\frac{m}{b_0\,c_1}}\,(c_1 r + c_2)\right]. \tag{49}$$

where $c_2$ is another constant. Note that the vacuum condition $r \to \infty$ is satisfied in such way that $\beta_\theta \to \sqrt{\frac{b_0\,c_1}{m}} = b_\theta = constant$. So the norm of the Bumblebee field $\beta^\mu(r)$ is given by

$$|\beta(r)|^2 = (\beta_0(r))^2 - (\beta_r(r))^2 - (\beta_\theta(r))^2 = b_0^2 - \frac{b_0\,c_1}{m}\tanh^2\left[\sqrt{\frac{m}{b_0\,c_1}}\,(c_1 r + c_2)\right], \tag{50}$$

that we express this solution in terms of the vacuum parameters, which results in

$$|\beta(r)|^2 = b_0^2 - b_\theta^2\tanh^2\left[\frac{b_\theta}{b_0}mr + \frac{c_2}{b_\theta}\right]. \tag{51}$$

Without loss of generality, we assume at the vacuum limit that $b_0 = qb_\theta$, where $q$ is a positive real number and $c_2/b_\theta = -mr_0$, so

$$r_0 = -\frac{1}{mb_\theta}c_2, \tag{52}$$

where $r_0$ is a fixed position, such that eq. (51) can be rewritten as

$$|\beta(r)|^2 = b_\theta^2\left\{q^2 - \tanh^2\left[\frac{m}{q}(r - r_0)\right]\right\}. \tag{53}$$

---

[2] We remark that the choice $\left(\frac{d\beta_\theta}{dr} - \frac{m}{b_0}\beta_\theta^2\right) = +c_1$ implies that $\beta_\theta$ becomes proportional to the tangent what makes this component to diverge.

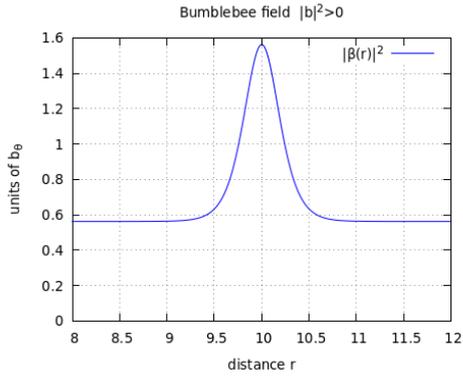

FIG. 1: Bumblebee field time-like for $q = 1.25$.

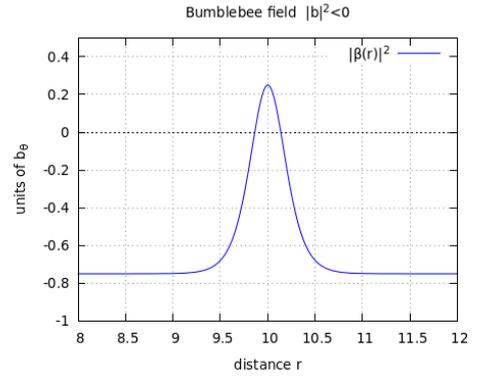

FIG. 2: Bumblebee field space-like for $q = 0.5$.

At the vacuum limit and the asymptotic condition $r \to \infty$, we have that

$$|\beta(r)|^2 \to b^2 = b_\theta^2(q^2 - 1) = b_0^2 - b_\theta^2. \tag{54}$$

So if $b_0 > b_\theta$ the Bumblebee vector field at the vacuum limit it is a time-like vector, or $|\beta(r)|^2 = b^2 > 0$, and if $b_\theta > b_0$ then the Bumblebee vector field at the vacuum limit is a space-like vector with $b^2 < 0$. Moreover $\beta(r)$ modulates the magnetic field (43). Then if we take $q^2 = 1 + \zeta$ where $\zeta > 0$ concerning to Bumblebee time-like vector, and $q^2 = 1 - \zeta$ where $0 < \zeta < 1$ concerning to Bumblebee space-like vector, the equation (53) can be write as

$$|\beta(r)|^2 = b_\theta^2 \left\{ \pm\zeta + \mathrm{sech}^2 \left[ m(r - r_0) \right] \right\}, \tag{55}$$

which corresponds to a soliton behaviour close to $r_0$ as shown in Fig.(1) for $q = 1.25$, in the time-like sector, and in Fig.(2) for $q = 0.5$ in the space-like sector. Moreover, note that this Bumblebee soliton explicitly depends on the topological mass $m$. We stress that the topological mass vanishes at $r \to \infty$, and therefore there is no Bumblebee soliton solution at this asymptotic condition.

In the core of the vortex, we obtain the vector potential $A(r)$ through the equations (34), (35) and (36). Indeed, by taking the condition $\beta_r(r) = 0$, we eliminate Eq.(35) and, consequently, we combine Eqs. (34) and (36), in such a way that

$$\frac{d^2\beta_\theta(r)}{dr^2} + \left( \frac{1}{r} - 2m \right) \frac{d\beta_\theta(r)}{dr} = \left[ \left( \frac{n}{r} - eA(r) \right)^2 + \alpha^2 + 2\lambda|\beta(r)|^2 \right] (\beta_\theta + \beta_0(r)) , \tag{56}$$

where $\beta_\theta$ and $|\beta(r)|^2$ are given, respectively, by (49) and (51). Thus, considering $\frac{1}{r} < m$, we find that,

$$A(r) = \frac{n}{er} - \frac{1}{e} \left[ \frac{\frac{m}{q} \left( \frac{2m}{q} \tanh[m(r - r_0)] + 2m \right) \mathrm{sech}^2[m(r - r_0)]}{q - \tanh[m(r - r_0)]} - \alpha^2 - 2\lambda b_\theta^2 \left( q^2 - \tanh^2[m(r - r_0)] \right) \right]^{1/2}, \tag{57}$$

which is an estimative for the vector potential near to $r_0$ in the vortex core, where we have considered $q = b_0/b_\theta$ with $q \neq 1$. Here one can obtain the magnetic field from potential vector $A(r)$ employing Eq. (38).

The behaviour of the potential vector $A(r)$ in the vortex core of the time-like sector of the Bumblebee field is shown in the Fig. (3). Note that, as the topological mass $m$ increases, $A(r)$ becomes more localized, and the magnitude of the magnetic field $H(r)$ increases, as it is shown in Fig.(4).

The behaviour of the fields $A(r)$ and $H(r)$ in the vortex core of the Bumblebee field in the space-like sector are shown in the Fig. (5) and Fig. (6), respectively. In this case, the position $r_0$ seems to be a limit for the propagation of the Bumblebee field, that is, the Bumblebee field is free to propagate in $0 < r < r_0$, however, at $r_0$ limit the magnetic field diverges.

Furthermore, we observe that the Bumblebee field becomes constant to $r \to \infty$ and satisfies the relation (30), $b^2 = b_0^2 - b_\theta^2$ and $A(r) \to \frac{n}{er}$. To the asymptotic condition $r \to \infty$ we get $\tanh[m(r - r_0)] \to 1$, $\mathrm{sech}[m(r - r_0)] \to 0$, and $b^2 = -\alpha^2/2\lambda$, which leads eq. (57) to

$$A(r) = \frac{n}{er} - \frac{1}{e} \left[ -\alpha^2 - 2\lambda \left( b_0^2 - b_\theta^2 \right) \right]^{1/2} = \frac{n}{er} . \tag{58}$$

As expected, it corresponds to a finite energy vortex. Although we remark that eq. (57) shows that the potential vector $A(r)$ has a non-trivial behaviour in the vortex core, indeed it explicitly depends on the topological mass $m$, except for $r \to \infty$ vacuum limit.

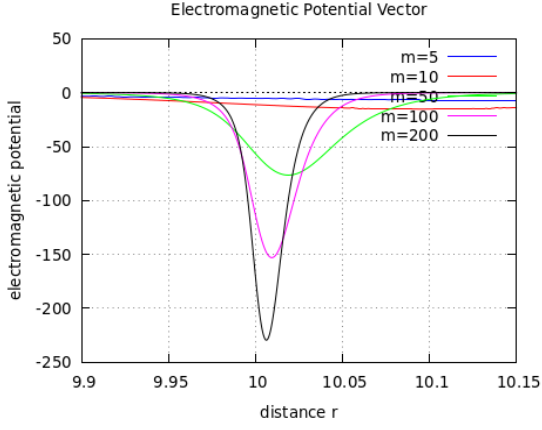

FIG. 3: Potential vector at $r_0$ for Bumblebee time-like with several values of topological mass $m$. We use $q = 1.25$, $b_\theta = 1$, $e = 1$, $\lambda = 2$, $r_0 = 10$, and $\alpha^2 = -2.25$.

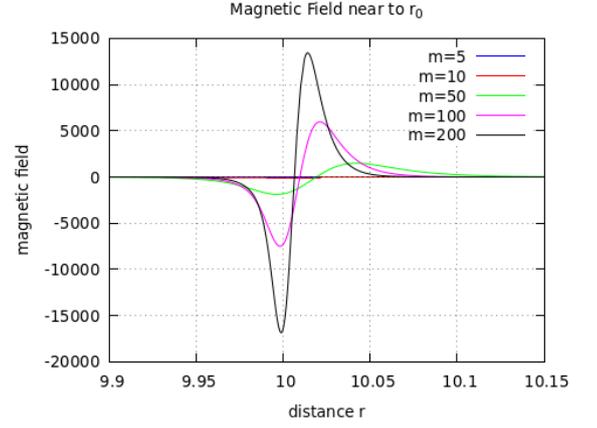

FIG. 4: Magnetic field at $r_0$ for Bumblebee time-like with several values of topological mass $m$. We use $q = 0.5$, $b_\theta = 1$, $e = 1$, $r_0 = 10$, $\lambda = 2$, and $\alpha^2 = 3$.

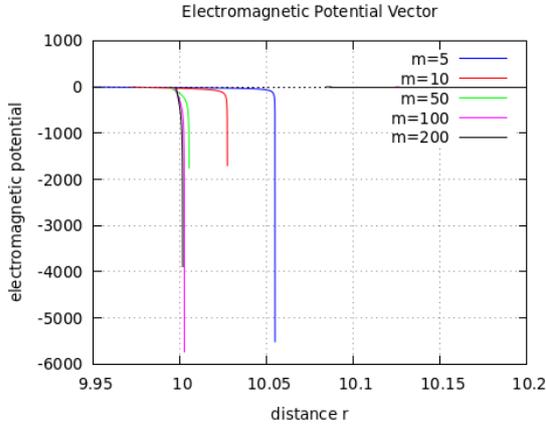

FIG. 5: Potential electromagnetic field near to $r_0$ for Bumblebee space-like with several values of topological mass $m$.

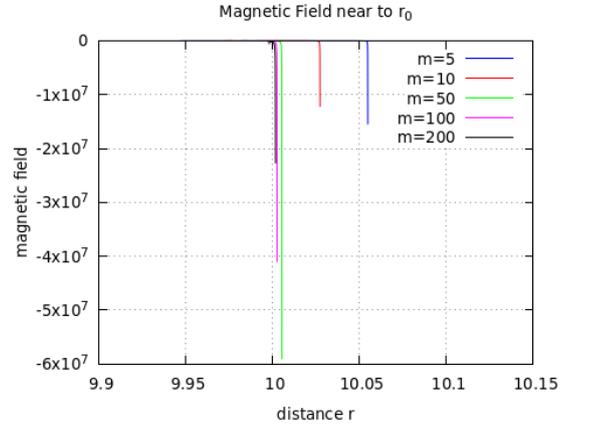

FIG. 6: Magnetic field near to $r_0$ for Bumblebee space-like with several values of topological mass $m$.

## IV. THE SPECTRUM ANALYSIS AND THE ASYMPTOTIC LIMIT BEHAVIOUR

Taking the Lagrangian (8), the propagators can be written down in three sectors, as

$$\Omega^{\mu\nu} = \frac{\partial^\mu \partial^\nu}{\Box} \quad , \quad \Theta^{\mu\nu} = \eta^{\mu\nu} - \frac{\partial^\mu \partial^\nu}{\Box} \quad , \quad S^{\mu\nu} = \epsilon^{\mu\lambda\nu}\partial_\lambda \,, \tag{59}$$

where $\Omega$ is the the longitudinal, $\Theta$ and $S$ are the transversal sectors of the propagator remarking that $S$ has topological origin. Thereby the invertible operator is $\mathcal{O}^{\mu\nu} = 2mS^{\mu\nu} + (\Box + \alpha^2)\eta^{\mu\nu}$, and its inverse $\mathcal{O}^{-1}$ is

$$(\mathcal{O}^{\mu\nu})^{-1} = -\frac{1}{(\Box + \alpha^2)}\,\Omega_{\mu\nu} + \frac{\Box + \alpha^2}{(\Box + \alpha^2)^2 + 4m^2\Box}\,\Theta_{\mu\nu} - \frac{2m}{(\Box + \alpha^2)^2 + 4m^2\Box}\,S_{\mu\nu}. \tag{60}$$

Converting to the momenta space, the $B$-field propagator can be written down as ,

$$\langle B^\mu B^\nu \rangle_L(k) \;=\; -\frac{i}{k^2 - \alpha^2}\left(\frac{k^\mu k^\nu}{k^2}\right), \tag{61}$$

$$\langle B^\mu B^\nu \rangle_T(k) \;=\; \frac{i}{(k^2 - \alpha^2)^2 - 4m^2 k^2}\left[(k^2 - \alpha^2)\left(\eta^{\mu\nu} - \frac{k^\mu k^\nu}{k^2}\right) + 2mi\epsilon^{\mu\lambda\nu}k_\lambda\right], \tag{62}$$

where we have separated in longitudinal ($L$) and transversal ($T$) sectors. Observe that the longitudinal sector exhibits a simple pole $k^2 = \alpha^2$, and if we saturate this sector with the conserved current it vanishes, thereby it has no dynamics and therefore we can throw it away. On the other hand the transversal sector can be re-written in such a way that,

$$\langle B^\mu B^\nu \rangle_T(k) = \frac{i}{(k^2 - \rho_+^2)(k^2 - \rho_-^2)} \Big[ (k^2 - \alpha^2)(\eta^{\mu\nu} - \frac{k^\mu k^\nu}{k^2}) + 2mi\epsilon^{\mu\lambda\nu}k_\lambda \Big], \tag{63}$$

where the massive poles are,

$$\rho_\pm^2 = \alpha^2 + 2m^2 \pm 2m\sqrt{\alpha^2 + m^2}. \tag{64}$$

We note that the transversal sector poles represent the unique dynamic degrees of freedom of the model, and if we take $m = 0$ we get the usual Proca-type massive pole as expected. The important point here is that we can easily observe that the topological mass contributes to dynamic mass poles $\rho_\pm^2$. And to obtain such contribution we suggest a simple relation between usual mass and the topological mass terms, that is $\varepsilon = \frac{\alpha}{m}$, which means that usual mass term $\alpha$ is a multiple of the topological mass $m$ term, and $\varepsilon$ is a constant value. Note that $\alpha$ is a non-zero value considering the initial condition of the vacuum symmetry break given by potential Eq. (3). In this way we can substitute it in (64), giving $\rho_\pm^2 = C(\varepsilon)m^2$, where $C(\varepsilon)$ is the constant expression $C(\varepsilon) = \varepsilon^2 + 2 \pm 2(1 + \varepsilon^2)^{1/2}$. We observe that for real values $C(\varepsilon)$ is always positive. So, in spite that the Chern-Simons-type sector does not contribute to the vacuum, the topological mass term $m^2$ yields the "speed" of the field to arrive at the asymptotic limit, and the size of the region where $H$, that is different from zero on the vortex core. So, in order to include a Chern-Simons-type contribution to the mass pole $\rho$ we have to "deform" the topological mass $\textsc{m}$ in the above $C(\varepsilon)$ in such way that

$$\textsc{m}_\pm^2 \to C(\varepsilon)m^2 \equiv \rho_\pm^2 \tag{65}$$

for $C(\varepsilon) > 0$, where we can rewrite the Bumblebee field as

$$|\beta(r)| = b_\theta \operatorname{sech}\left[\textsc{m}_\pm(r - r_0)\right]. \tag{66}$$

Then the Bumblebee solitons, or vortices, can have two positive modes ($\pm$) that depend on the mass deformation factor $C(\varepsilon)$.

## V. CONCLUSION

In this work we have analysed the contribution of the vacuum and the topological mass to the asymptotic behaviour of a Bumblebee field in $(1 + 2)D$. To this goal we have recalled the Bumblebee vector field originated from the string model which represents the Lorentz violation effect observed by Kostelecky and Samuel [3]. Based on these works and from the Maxwell-Chern-Simons[1, 38] we have constructed a model to obtain the topological elements of the Bumblebee model as long as the the vacuum Lorentz violation parameter.

In spite of a non-exact and/or non-analytic solution of the differential equations which were derived from each component of the equation of motion, the topological characteristic of the asymptotic solutions [1, 2, 35] are obtained and are shown in the equations(33), (34) and (35), (36). The resulting topological objects can be interpreted as vortices [35], where each one of their components are shown to be a polarized soliton in the proper space-time direction. We have also shown that a possible origin for these vortices comes from the non-trivial vacuum of the Bumblebee field $B_\mu$ dynamics.

In more details, we have obtained that the vortex originated from the Bumblebee field with polarized fixed direction $b_\mu$ can be associated with a anisotropy of vacuum energy (53) that is localized at a fix point $r_0$. This fixed point strongly depends on the inverse of product of the module of the Bumblebee field and the topological mass (52). Recalling the studies on Lorentz symmetry violation studies [22–27], there is an interesting similarity where the (local) Lorentz symmetry violation might induce a local space-time perturbation giving a dynamical anisotropic anomaly that is manifest by a vortex object. It suggests that the possible appearance of topological vortices at fixed point $r_0$ in the limit of very high energy physics indicates that we have start off from a model where Lorentz violation terms participate in the dynamics. We have also obtained that if the Bumblebee field is a time-like vector the magnetic appears as a "pulse" centred in $r_0$ in a limitless universe sketched in Fig.(4) , while if the Bumblebee field is a space-like vector the magnetic appears as a barrier limiting the universe sketched in Fig.(6). The last one observation is interesting because the Lorentz violation yields a kind of boundary to the dynamics which classically can be interpreted as a symmetry violation in a orthogonal direction. It remains to obtain the physics value of the $r_0$, that can be determine in the quantum regime, which is the issue of a forthcoming work.

Furthermore, for the sake of completeness, we have analysed the role played by the topological Chern-Simons-type terms. Although they do not contribute to the non-trivial vacuum obtained, by means of a mass pole verification using propagators computation we observe that it can contribute with a positive two-mode-extension to the mass term, which indicates that its factor preserves the topological feature of the asymptotic limit of the field. It can characterize the size of the region where the "magnetic" field is non-zero, or core vortex, and the "speed" of the field to saturate the asymptotic limit. In the scope of the Quantum Field Theory it could be noted that topological vortices stem from a possible optical vacuum energy state or dynamical topological defect. Thus these objects might be present in models that due to the space structure have anisotropic (optical or not) effects.

## VI. ACKNOWLEDGEMENTS

LPC would like to thank to Prof. J.A. Helayl-Neto for discussions and the kind hospitality at the CBPF.

------------------------------------------------